\documentclass[aps, prl, reprint, superscriptaddress, floatfix]{revtex4-1}
\usepackage{amsmath}
\usepackage{amsfonts}
\usepackage{graphicx}
\usepackage{cancel}
\usepackage{hyperref}
\usepackage{color}

\newcommand{\order}[1]{\mathcal{O}\!\left(#1\right)}

\begin{document}
	
	\title{Lindbladian quantization of mechanical systems with nonholonomic constraints}
	\author{Daniel Schubring}
\affiliation{Physics Department, City College of the City University of New York, New York 10031, USA}
\author{Sriram Ganeshan}
\affiliation{Physics Department, City College of the City University of New York, New York 10031, USA}
\affiliation{Physics Program, Graduate Center of City University of New York, New York 10031, USA}
\begin{abstract}

Nonholonomic mechanics describes systems subject to non-integrable velocity constraints, such as
rolling bodies and skating motion.  These systems generally lack a canonical Hamiltonian formulation, obstructing standard quantization methods.  Here we quantize nonholonomic systems as Markovian open quantum systems, with the nonholonomic constraint appearing in a large-dissipation limit. We find explicit Lindblad superoperators that reproduce the classical dynamics of the Chaplygin sleigh and the Suslov problem in the semiclassical limit. The master equation is numerically simulated, and the covariance is shown to satisfy a relation predicted by the theory of metastability in open quantum systems.

\end{abstract}
	
	\maketitle

    \section{Introduction}
    Nonholonomic mechanics was developed in the late nineteenth and early twentieth centuries to describe rolling bodies and other mechanical systems subject to velocity constraints~\cite{Chaplygin1903Sphere, hertz1910prinzipien, Chaplygin1912ReducingMultiplier}. It remains central to geometric mechanics and control theory~\cite{Bloch2003Nonholonomic,Borisov2017Survey}, with direct implications for the design and control of mobile robots and autonomous vehicles. Beyond finite-dimensional mechanical systems,  nonholonomic structures have also been identified in continuum and fluid systems~\cite{bloch2025infinite, Monteiro2023BrokenParity, nabizadeh2025fluid, abanov2026infinite}.  Unlike holonomic constraints, which restrict the configuration space itself, nonholonomic constraints restrict the \emph{allowed velocities} through non-integrable relations. As a result, the equations of motion for genuinely nonholonomic systems generally cannot be obtained from Hamilton's principle with arbitrary variations. This absence of a conventional action principle complicates the construction of a canonical Hamiltonian formulation, since the natural ``almost-Poisson brackets" associated with nonholonomic dynamics typically violate the Jacobi identity~\cite{van1994hamiltonian,deLeon2024NewPerspective}. Consequently, conventional quantization schemes do not apply directly in this case, in contrast with holonomic constraints, for which Dirac's constrained Hamiltonian quantization is built on a well-defined canonical phase space and constraint algebra~\cite{dirac1950generalized,dirac2013lectures}. 
    
    Several approaches to nonholonomic quantization have been proposed~\cite{Eden1951Hamiltonian,Eden1952Quantum,deLeon2024NewPerspective,Bloch2008Quantization,Fernandez2018Nanocar,Fernandez2022Chaplygin}, but there is still no settled, generally applicable prescription. The question of quantizing nonholonomic systems is also of practical relevance in view of the rapid development of molecular machines, including molecular cars, wheels, and synthetic motors~\cite{Shirai2005Nanocars,Grill2007MolecularWheel,ErbasCakmak2015MolecularMachines}, where quantum effects could interplay with nonholonomic constraints. A further motivation for studying quantum nonholonomic constraints is to understand the kinds of (non) equilibrium quantum dynamics that can arise from them. Classical nonholonomic systems occupy an unusual position between equilibrium and non-equilibrium dynamics: the ideal constrained dynamics conserves energy, yet the constraint forces possess an ``intrinsic dissipative" nature exemplified best in the Chaplygin sleigh~\cite{Kozlov1983Realization,Kozlov1992Realizing, Bloch2003Nonholonomic}.  The general question of whether nonholonomic dynamics can be obtained as the limiting behavior of systems with friction was first raised by Carath\'eodory~\cite{Caratheodory1933, Eldering2016Friction}. It is now understood that, when the relevant friction forces are taken to be infinitely large, the nonholonomic equations emerge as a singular limit of the dissipative dynamics. Relatedly, classical effective-action descriptions in which a bath realizes dissipative and nonholonomic dynamics have recently been developed~\cite{Besharat2024Effective}.
    In this work, we use this observation as the starting point for a different route to quantization. Rather than attempting to quantize the nonholonomic bracket directly, we formulate the constraint problem as a Markovian open quantum system. The basic idea is to first realize the velocity constraint as the infinite-friction limit of a dissipative classical dynamics, and then quantize the corresponding finite-friction system using a completely positive Lindblad evolution~\cite{Gorini1976,Lindblad1976,BreuerPetruccione}. In this formulation, the nonholonomic constraint is not imposed as an operator identity on a reduced Hilbert space. Instead, it emerges dynamically from the separation between fast dissipative relaxation of the forbidden velocity components and slow evolution along the allowed directions. We explicitly demonstrate this construction for three closely related nonholonomic systems: the Chaplygin sleigh, its skater limit, and the Suslov problem or system. From the perspective of open quantum systems, the statement that an infinite-dissipation limit produces a constraint is naturally interpreted as the emergence of a metastable manifold of long-lived modes of the Lindblad generator~\cite{Macieszczak2016Metastability,Macieszczak2021Classical}. For the skater system, we analytically show how this metastable manifold emerges in the large-dissipation limit and verify the predicted covariance through numerical simulations of the master equation. 

\section{The Chaplygin sleigh and skater system} 
\begin{figure*}
    \centering
   \includegraphics[width=0.4\textwidth]{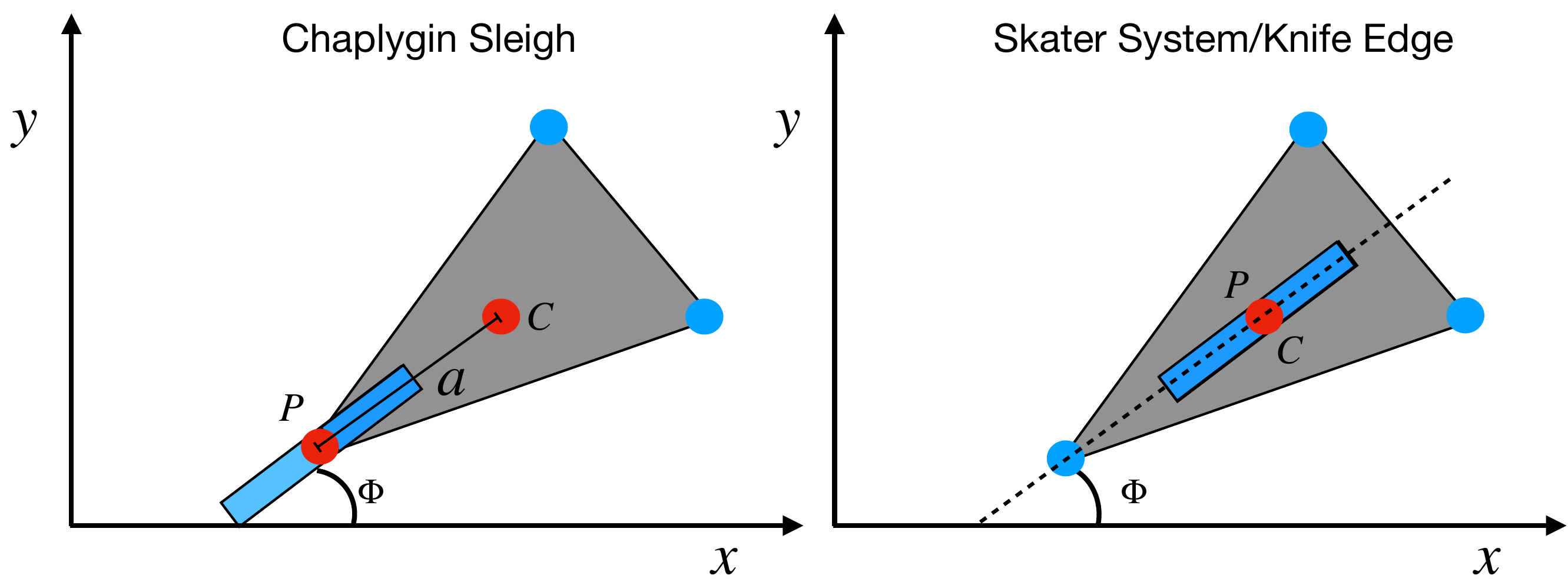} \includegraphics[width=0.5\textwidth]{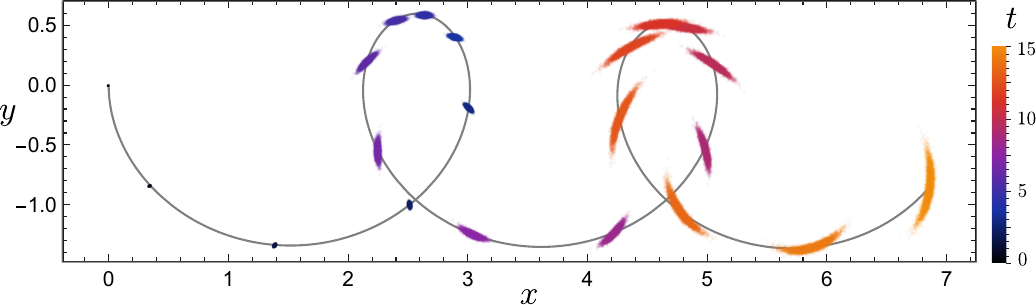}
    \caption{ Left: Schematic of the Chaplygin sleigh and skater geometry. The point $P$ is the skate contact point, $C$ is the center of mass, and the body orientation is $\Phi$ relative to the laboratory $(x,y)$ axes. For the Chaplygin sleigh, the center of mass is displaced from $P$ by the distance $a$; the skater limit corresponds to $a=0$. Right: Results of a simulation of the master equation \eqref{Master eq skater} of the skater system with dimensionless parameters $\gamma=10^2, \hbar=10^{-7}, \alpha=2/3$. The evolution of the quantum state as a function of time $t$ is represented by plotting 10,000 samples of the marginal probability distribution over $x,y$ at intervals of 75 time steps, with time indicated by color. The deterministic nonholonomic trajectory is plotted as a solid curve.}
    \label{fig:skaterXY}
\end{figure*}
We begin with the Chaplygin sleigh, a paradigmatic example of a nonholonomic mechanical system~\cite{Chaplygin1912ReducingMultiplier,Caratheodory1933}. A schematic of the setup is shown at left in Fig.~\ref{fig:skaterXY}. The Chaplygin sleigh consists of a rigid body moving in the plane with a knife edge, or skate, attached at a fixed point on the body. The skate constrains the motion by forbidding velocity transverse to its blade, while allowing motion along the blade direction. Since this restriction is imposed on the velocity rather than on the configuration itself, the Chaplygin sleigh provides a minimal and physically transparent setting in which to study nonholonomic dynamics. The configuration space is described by coordinates \(q^i=(x,y,\Phi)\), where \((x,y)\) denotes the position of the skate contact point in the plane and \(\Phi\) is the orientation angle of the body. The center of mass is displaced from the skate by a fixed distance \(a\) along the body frame. This offset distinguishes the Chaplygin sleigh from the simpler centered knife-edge, or skater, problem, and is responsible for the characteristic coupling between translation and rotation. The velocity $v_s$ perpendicular to the skate is constrained as
\begin{equation}
v_s =-\sin\Phi\,\dot x+\cos\Phi\,\dot y=0.
\label{eq:vs}
\end{equation}
Note that the above constraint is genuinely nonholonomic since it restricts the allowed velocities at each configuration but cannot be integrated into a constraint on the configuration variables alone. It will be convenient to introduce the rotated coordinates $r= \cos\Phi\,x + \sin \Phi\,y,\, s= -\sin \Phi\, x + \cos \Phi\, y,\, p_r=\cos\Phi\,p_x + \sin \Phi\,p_y,\,p_s= -\sin \Phi\, p_x + \cos \Phi\, p_y$. The Hamiltonian in these non-canonical coordinates is
\begin{equation}
	H=\frac{p_r^2+p_s^2}{2m}+\frac{\left(p_\Phi-ap_s\right)^2}{2I_0}+\alpha y.
\end{equation}
We include a linear potential, a standard choice in the $(a=0)$ limit, corresponding to the skater or knife-edge system~\cite{Kozlov1983Realization,Bloch2003Nonholonomic}. The equations of motion of the unconstrained dynamics are given by Hamilton's equations. The nonholonomic constraint in Eq.~\ref{eq:vs} is then implemented by adding a damping force $-2\gamma m v_s$ to the equation of motion for $p_s$. As $\gamma\to\infty$, this drives the forbidden velocity to scale as $v_s\sim \gamma^{-1}$, while the product $-2\gamma m v_s$ remains $\mathcal O(1)$. This finite limiting term plays the role of the constraint force in the standard Lagrange--d'Alembert formulation~\cite{Bloch2003Nonholonomic, Caratheodory1933, Eldering2016Friction, Kozlov1983Realization}.

To perform a Lindbladian quantization, we note that this damping term modifies the Liouville equation for a probability distribution $\rho$ on phase space to
\begin{align}
	\dot \rho=\mathcal{L}\rho\equiv -[\rho,H]_P+2\gamma\partial_{p_s}\left(m v_s \rho\right),
\label{Liouville Chaplygin}
\end{align}
where we use a subscript $P$ to indicate Poisson brackets, and reserve curly brackets for quantum mechanical anticommutators. We then quantize this dissipative Liouville dynamics using a completely positive Lindblad master equation \cite{Gorini1976,Lindblad1976,BreuerPetruccione}:
\begin{equation}
	\begin{split}
	\dot{\rho}=-\frac{1}{i\hbar}\left[\rho, H_\gamma\right]+\sum_a \gamma_a \mathcal{D}_a \rho,\\\mathcal{D}_a \rho \equiv  \frac{1}{\hbar}\left(L_a \rho L_a^\dagger - \frac{1}{2}\{L_a^\dagger L_a, \rho\}\right).\label{Master eq}
	\end{split}
\end{equation}
Here $\rho$ denotes the density matrix, and we will later use the same symbol for its Wigner transform. $H_\gamma$ is a Hamiltonian that may include $\gamma$-dependent counterterms.

Upon decomposing the Lindblad operators as $L=A+i B$ with $A, B$ Hermitian, we have
\begin{multline}
	\mathcal{D}_a\rho = \frac{1}{i\hbar}\left[A_a,\{\rho, B_a\}\right]+\frac{1}{i\hbar}\left[\rho, \frac{1}{2}\left\{A_a,B_a\right\}\right]\\
	-\frac{1}{2\hbar}\left(\left[A_a,\left[A_a, \rho\right]\right]+\left[B_a,\left[B_a, \rho\right]\right]\right).\label{Dissipator}
\end{multline}
The first term gives the \emph{drift} that survives the classical limit. The second term may be canceled by a counterterm in $H_\gamma$. The third term gives \emph{diffusion} terms, which are suppressed in the classical limit by a factor of $\hbar$.

To represent a constrained mechanical system, the operators $B$ will be chosen to involve the constrained velocity combinations ( for example $\sin\Phi\, \dot x\, \text{and}\, \cos \Phi\, \dot y$ for the Chaplygin sleigh), and the operators $A$ will be chosen to involve coordinates in configuration space. Then in the large $\gamma$ limit, the drift term will end up suppressing the constrained velocities after a short time $\sim\gamma^{-1}$. 

A simple set of Lindblad operators for the Chaplygin sleigh is
\begin{align}
	\frac{L_1}{\sqrt{D}} = x-\frac{im}{2D}\left\{\sin\Phi, v_s\right\},\quad \frac{L_2}{\sqrt{D}}=y+\frac{im}{2D}\left\{\cos\Phi, v_s\right\}.\label{Lindblad operators Chaplygin}
\end{align}
where the diffusion coefficient $D$ has dimensions of action divided by length squared. For simplicity, we now focus on the $a=0$ skater system, where $v_s = p_s/m$. The $a\neq 0$ case involves a kinetic metric with off-diagonal components, and it is deferred to Appendix~\ref{App chaplygin}.

These Lindblad operators generate a master equation for the density matrix $\rho$ as in \eqref{Master eq}. To compare with the equation for the probability distribution $\rho$ in \eqref{Liouville Chaplygin} we will do a Wigner transform, but even before considering the detailed transform, we may see how this choice of Lindblad operators \eqref{Lindblad operators Chaplygin} reduces to the classical result. At lowest order, we may replace anticommutators with direct multiplication and commutators with $i\hbar$ times a Poisson bracket. The drift terms in \eqref{Dissipator} have lowest-order transform,
\begin{equation}
\left[x, -\sin\Phi \,m v_s\,\rho\right]_P + \left[y, \cos\Phi  \,m v_s \,\rho\right]_P= \partial_{p_s}\left(m v_s \rho\right).\label{Classical drift}
\end{equation}
Thus the classical dissipative drift term is reproduced.  The set of operators \eqref{Lindblad operators Chaplygin} is not the unique choice reproducing the classical drift. As shown in Appendix~\ref{App general diffusion}, choices may yield different diffusion terms.

The quantization of the free Hamiltonian is standard, and of course the angular momentum $p_\Phi$ is quantized in units of $\hbar$. The exact Wigner transform of the master equation will also involve the Wigner function $\rho$ depending on a discrete variable $p_\Phi$. But we may make a continuum approximation which is valid if the variance in $p_\Phi$ is sufficiently large, 
$\sigma_{p_\Phi}^2\gg \hbar^2.$

Under this approximation, the Wigner-transformed master equation becomes of the Fokker-Planck type,
\begin{gather}
\dot{\rho}=\mathcal{L}\rho+\frac{\hbar\gamma D}{2}\left(\partial_{p_r}^2+\partial_{p_s}^2\right)\rho+\frac{\hbar\gamma}{2D}\mathcal{Q}\rho,\label{Master eq skater}\\
\mathcal{Q}\rho \equiv \left(\partial_s^2+2 p_r \partial_s \partial_{p_\Phi}+\left(p_r^2+p_s^2\right)\partial_{p_\Phi}^2\right)\rho.\label{Diff term}
\end{gather}
The full derivation is given in Appendix~\ref{App wigner}, but this Fokker-Planck approximation can be verified by starting with the operator form of the master equation \eqref{Master eq} and replacing commutators by Poisson brackets as above.

It is convenient to switch to a system of units with $m=I_0=D=1$, and understand $\hbar, \gamma$ as the dimensionless parameters $m\hbar/(I_0D),  m\gamma/D$. A minimum uncertainty wave packet in $\Phi$ and $p_\Phi$ will have variances $\sigma_\Phi^2\sim \hbar, \sigma_{p_\Phi}^2\sim \hbar$. So if $\hbar \ll 1$, the packet will be highly localized but still satisfy the criterion $\sigma_{p_\Phi}^2\gg \hbar^2$ necessary for the continuum approximation.


Equation \eqref{Master eq skater} may be simulated by sampling trajectories of an associated Langevin equation (see e.g. \cite{Gardiner1985}). Such a simulation beginning from a minimum uncertainty wave packet is shown in Fig.~\ref{fig:skaterXY}. The marginal distribution of $\rho$ in the $x, y$ plane is plotted, and the wave packet closely follows the classical nonholonomic trajectory.

\section{The Suslov system}

        The Suslov system (see e.g. \cite{Fedorov1995RigidBody}) is a closely related system to the Chaplygin sleigh that involves rigid-body rotations that are constrained about a left-invariant (body-frame) axis. For this case, instead of a body moving in the plane with a knife edge, we consider a rigid body whose angular velocity is not allowed to have a component along one fixed body-frame axis. The phase space for rigid body rotations is described in terms of Euler angles and body-frame angular momentum $j_a$. Left-invariant vector fields $X_a$ are used to take angular derivatives. 	It will be convenient to describe the angular position with an $SO(3)$ matrix $R$, which satisfies
		\begin{align}
			X_a\left(R_{bc}\right)=\epsilon_{acd}R_{bd}.
		\end{align}	
		A concrete description of $X$ and $R$ in terms of Euler angles is given in Appendix~\ref{App suslov}. The associated Poisson brackets are given by
		\begin{align}
		[f,g]_P=X_a(f)\partial_{j_a}g-X_a(g)\partial_{j_a}f-j_a \epsilon_{abc}\partial_{j_b}f\partial_{j_c}g.\label{Poisson brackets Suslov}
	\end{align}
	The equations of motion are given by taking Poisson brackets with the Hamiltonian $H= \frac{1}{2}j_aI^{-1}_{ab}j_b$, where $I$ is the moment of inertia tensor. There is a nonholonomic constraint on the body-frame velocity $\omega_1 = I^{-1}_{1a}j_a\sim 0$. Implementing this with dissipation as before, the classical drift part of the Fokker-Planck equation becomes
\begin{align}
	\dot{\rho}=\mathcal{L}_{SO(3)}\rho\equiv-\left[\rho, H\right]_P + 2\gamma\partial_{j_1}\left(\mu\, \omega_1 \rho\right),\label{Liouville Suslov}
\end{align}		
	where $\mu$ has dimensions of moment of inertia.
	
The dissipative part of the drift term in \eqref{Liouville Suslov} is reproduced by the set of Lindblad operators
\begin{align}
L_a = \sqrt{\kappa}R_{a3}-i\frac{\mu}{2\sqrt{\kappa}}\{R_{a2}, \omega_1\}.\label{Lindblad Suslov}
\end{align}
The diffusion coefficient $\kappa$ has units of action. These operators are structurally similar but distinct from previously appearing operators for the thermalization of quantum rotors \cite{Stickler2018Rotational}.

The operator form of the master equation is again given by \eqref{Master eq} and \eqref{Dissipator}. A full Wigner transform may be carried out along the lines of \cite{Mukunda2004WignerLieGroups}. Here we again work in the Fokker-Planck approximation, replacing commutators in the master equation by $i\hbar$ times the Poisson brackets \eqref{Poisson brackets Suslov},
\begin{gather}
\dot{\rho}=\mathcal{L}_{SO(3)}\rho+\frac{\hbar \gamma \kappa}{2}\left(\partial_{j_1}^2+\partial_{j_2}^2\right)\rho + \frac{\hbar\gamma \mu^2}{2\kappa}\mathcal{Q}\rho.\label{Master equation Suslov}
\end{gather}
The operator $\mathcal{Q}$ involves second derivatives with respect to $j$ and arises from the double commutator of $\{R_{a2},\omega_1\}$. It is shown in detail in Appendix~\ref{App suslov}.

Note that while the Fokker-Planck equation for the $a\neq 0$ Chaplygin sleigh may be found directly from the Lindblad operators \eqref{Lindblad operators Chaplygin}, it is also a limiting case of the equation for Suslov \eqref{Master equation Suslov} with a particular moment of inertia tensor. More details on the construction are in Appendix~\ref{App chaplygin}.

\begin{figure}
    \centering
    \includegraphics[width=\columnwidth]{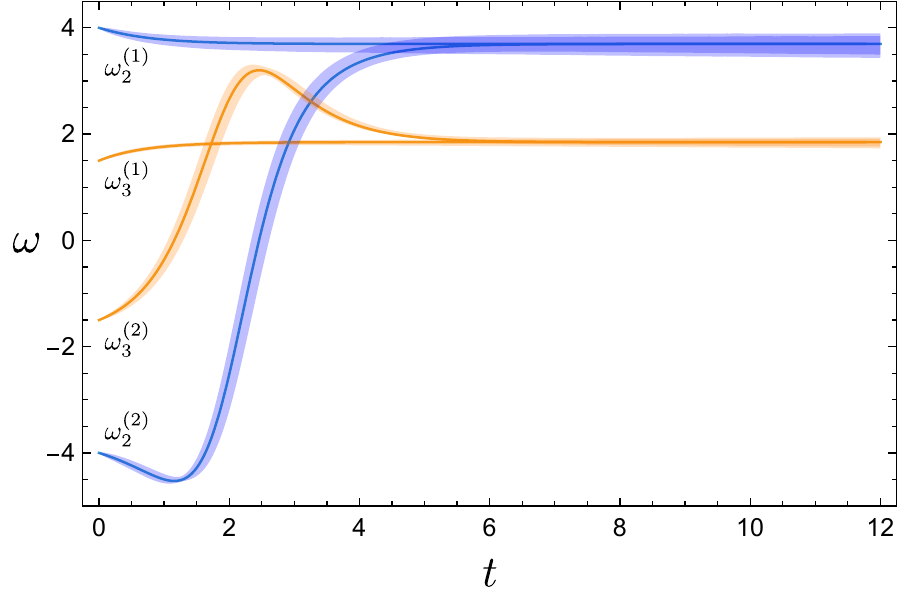}
    \caption{A plot of the unconstrained angular velocities $\omega_2$ (blue) and $\omega_3$ (orange) versus $t$ in the Suslov system. The deterministic nonholonomic result is given by the solid lines. Two different choices of initial conditions labeled by superscripts $(1)$ and $(2)$ converge to the same limiting state. The master equation \eqref{Master equation Suslov} is simulated with $\gamma=10^3, \hbar=10^{-6}$, and the shaded region represents values of $\omega$ within one standard deviation of the mean of the wave packet. }
    \label{fig:suslovOmega}
\end{figure}

Both the Suslov and $a\neq 0$ Chaplygin sleigh systems classically involve convergence of the momentum to a stable limit. To illustrate similar behavior in the quantum system, we carried out a Langevin simulation of equation \eqref{Master equation Suslov} in Fig.~\ref{fig:suslovOmega}.

\section{The metastable manifold} 
The examples above point to a common structure: the Lindblad dynamics rapidly damps the velocity component forbidden by the nonholonomic constraint, while the remaining variables evolve on a slower effective constrained sector. This suggests a general picture in which the nonholonomic constraint is realized as a long-lived manifold of states selected dynamically by the open-system evolution. This emergent scale separation in the large $\gamma$ limit can be elucidated by considering the theory of metastability in quantum open systems \cite{Macieszczak2016Metastability,Macieszczak2021Classical}. To illustrate this approach, we will focus on the momentum variables in the skater system with no potential ($a=0, \alpha=0$). The master equation is integrated over $r,s,\Phi$, and the marginal distribution is denoted by $f$,
	\begin{multline}
	\dot{f} = -\partial_{p_r}\left(p_\Phi p_s f\right)+\partial_{p_s}\left(\left(p_\Phi p_r+2\gamma p_s\right) f\right)\\+\frac{\hbar\gamma}{2}\left(\partial_{p_r}^2+\partial_{p_s}^2 +\left(p_r^2+p_s^2\right)\partial_{p_\Phi}^2\right)f.\label{Master eq skater f}
\end{multline}
Now we expand about a classical solution to the equations of motion $\bar{p}_{r,s}(t)$ with $p_\Phi=\omega$ constant. $f$ is taken to depend on $t$ and the fluctuation variables $q$
\begin{align}
	q_{r,s}=\frac{p_{r,s}-\bar{p}_{r,s}(t)}{\sqrt{\hbar\gamma}}, \quad q_{\Phi}=\frac{p_{\Phi}-\omega}{\sqrt{\hbar\gamma}}.
\end{align}
The master equation becomes
	\begin{multline}
	\dot{f}=-\partial_{q_r}\left(\left(\omega q_s+\bar{p}_s q_\Phi\right) f\right)+\partial_{q_s}\left(\left(\omega q_r+\bar{p}_r q_\Phi +2\gamma q_s\right) f\right)\\+\frac{1}{2}\left(\partial_{q_r}^2+\partial_{q_s}^2 +\left(\bar{p}_r^2+\bar{p}_s^2\right)\partial_{q_\Phi}^2\right)f+\order{\sqrt{\hbar\gamma}}.\label{Master eq weak noise}
\end{multline} 
If the higher-order terms are neglected, this master equation is exactly solvable, see Appendix~\ref{App ornstein uhlenbeck}. Even without the full solution, it can be determined that the stationary state has variance $\langle q_s^2\rangle\sim \gamma^{-1}$ and $\langle q_r^2\rangle, \langle q_\Phi^2\rangle \sim \gamma$. This suggests a rescaling of the variables
	\begin{align}
	q_s'= \gamma^{1/2}{q}_s,\quad q_r'= \gamma^{-1/2}{q}_r,\quad q_\Phi'= \gamma^{-1/2}{q}_\Phi,
\end{align}
and also $t'=\gamma^{-1}t$, $\bar{p}_s' = \gamma \bar{p}_s$ to eliminate $\gamma$ from the classical solution at leading order.

Now the master equation takes the schematic form $\gamma^{-1}\dot{f}=\left(\gamma\hat{H}_0  + \gamma^{-1}\hat{H}_1 +\gamma^{-3}\hat{H}_2\right)f$, with in particular
\begin{align}
	\hat{H}_0 f \equiv \partial_{q_s'}\left(\left(\omega q_r'+\bar{p}_r q_\Phi' +2 q_s'\right) f\right)+\frac{1}{2}\partial_{q_s'}^2f.
\end{align}

The large $\gamma$ limit may be understood through perturbation theory in $\gamma^{-1}$, and at zeroth order this involves the instantaneous eigenstates of $\hat{H}_0$. The kernel of  $\hat{H}_0$ is given by states of the form
\begin{align}
	f_{ms}= g(q_r', q_\Phi')\exp\left[{-2\left(q_s'+\frac{\omega q_r'}{2}+\frac{\bar{p}_r q_\Phi'}{2}\right)^2}\right],\label{Metastable state}
\end{align}
with $g$ an arbitrary profile function. States of the form $f_{ms}$ are stationary states at leading order, but when the perturbation $\gamma^{-1}\hat{H}_1$ is taken into account they pick up an eigenvalue correction of order $\sim \gamma^{-1}$ and are thus seen to be metastable, decaying on the (unprimed) time scale $t_{ms}\sim \gamma/\omega^2$.

In contrast, the higher-order eigenstates of $\gamma\hat{H}_0$ have large negative eigenvalues of order $\sim \gamma$ and they decay on the short time scale $t_{diss} \sim \gamma^{-1}$. So we may understand the meaning of the quantum nonholonomic constraint to be the restriction to the `metastable manifold' \cite{Macieszczak2016Metastability} of valid density matrices which are in the kernel of the leading-order terms $\gamma\hat{H}_0$ in the master equation.

\begin{figure}
    \centering
    \includegraphics[width=\columnwidth]{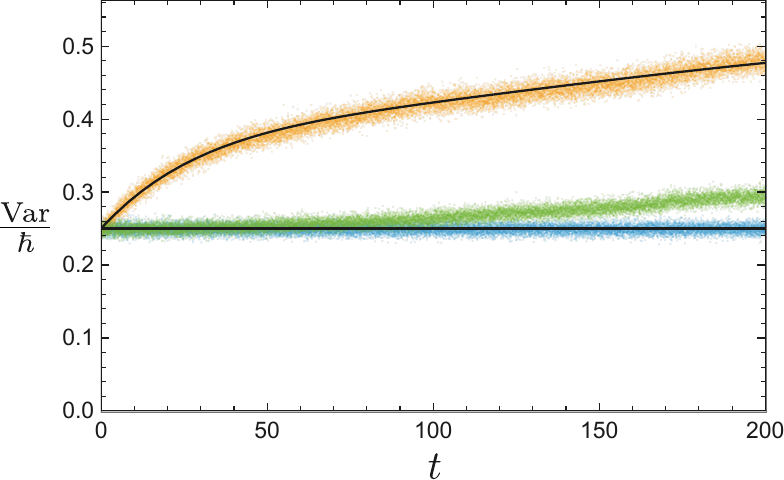}
    \caption{Variance of the constrained momentum $p_s$  (orange) versus time in the $\alpha=0$ skater system ($\gamma=100, \hbar=10^{-7})$. The mixed variance involved in \eqref{Variance nonholonomic} is plotted for $\hbar=10^{-7}$ (blue) and $\hbar=10^{-5}$ (green). Solid curves are calculated from a weak noise approximation and plotted points are sample variances from 4000 trajectories of a Langevin simulation.}
    \label{fig:skaterVar}
\end{figure}

In our case this means we are restricted to states with marginal distribution $f_{ms}$ \eqref{Metastable state}. In our original variables, this implies the nontrivial relation for the variance
\begin{align}
\text{Var}\left(p_s+\frac{\omega p_r}{2\gamma}+\frac{\bar{p}_r(t) p_\Phi}{2\gamma}\right)\approx\frac{\hbar}{4}.\label{Variance nonholonomic}
\end{align}
This result is supported numerically even for late times on the order of $t_{ms}$, as illustrated by the blue plotted points of Fig.~\ref{fig:skaterVar}. For later times or larger values of $\hbar\gamma$ the result is expected to depart from \eqref{Variance nonholonomic} due to the higher-order terms in \eqref{Master eq weak noise}, as illustrated by the green plotted points.

Note that while a number of approximations have been made to reduce the master equation of the skater system to a form in which the metastable manifold approach may be applied quantitatively, there is a closely related exactly solvable quantum open system. In the special case of the trivial classical solution $\bar{p}_{r,s}=0$, the master equation \eqref{Master eq weak noise} is equivalent to that of a damped harmonic oscillator with canonically conjugate position $\xi=q_r$ and momentum $\pi=q_s$, and Hamiltonian $\frac{\omega}{2}\left(\pi^2+\xi^2\right)$. Dissipation is implemented with a Lindblad operator $L=\xi+i\pi$ and the requisite Hamiltonian renormalization. Classically, the large $\gamma$ limit leads to the holonomic constraint $\pi=0$ with $\xi$ fixed at a constant value. The degeneracy of the metastable manifold corresponds to the classical degeneracy in the choice of position $\xi$.

\section {Discussion and Outlook}
		
	In this work, we have formulated a Lindbladian route to the quantization of the Chaplygin sleigh, the skater, and the Suslov system, which are paradigmatic examples of mechanical systems with nonholonomic constraints.  The starting point is not a reduced nonholonomic bracket, but rather a finite-friction realization of the constraint. In the infinite-friction limit, this dissipative dynamics reproduces the Lagrange--d'Alembert equations~\cite{Bloch2003Nonholonomic,Caratheodory1933,Eldering2016Friction,Kozlov1983Realization}, while at finite friction it admits a direct Markovian quantization by completely positive Lindblad evolution~\cite{Gorini1976,Lindblad1976,BreuerPetruccione}. The Chaplygin sleigh, the skater, and the Suslov system exhibit the same basic structure: the Lindblad drift damps the forbidden velocity component, the semiclassical dynamics approaches the corresponding classical nonholonomic trajectory on intermediate timescales, and the fast dissipative sector fixes the quantum fluctuations transverse to the constraint. 

    These examples suggest a broader framework in which a nonholonomic constraint distribution is implemented dynamically, rather than imposed as an operator identity. For a general collection of velocity constraints, the analog of the constructions above should identify the forbidden momentum directions and choose Lindblad operators whose classical drift damps precisely those components.
    To isolate the features that are genuinely tied to nonholonomicity, one should also develop the corresponding Lindbladian quantization of holonomic, or configurational, constraints. This comparison lies somewhat beyond the scope of the present manuscript and will be addressed separately in future work. 

The metastable-manifold picture \cite{Macieszczak2016Metastability,Macieszczak2021Classical} provided a clear organizing principle for our approach. The large-dissipation generator separates into fast modes that suppress constraint-violating motion and slow modes that encode the effective constrained dynamics. However, to apply this approach quantitatively we made use of a weak-noise approximation. At finite $\hbar\gamma$, the same Lindblad noise that enforces the constraint also produces diffusion in the remaining degrees of freedom. In the case of the skater system \eqref{Master eq skater f} this is seen in the equation for the variance of $p_\Phi$,
\begin{align}
    \frac{d}{dt}\langle p_\Phi^2\rangle = \hbar\gamma \left\langle p_r^2+p_s^2\right\rangle.
\end{align}
Unless the variances in $p_r$ and $p_s$ go to zero, the system does not settle into a stationary state; instead, it continues to heat. As shown in Appendix~\ref{App general diffusion}, the diffusion term causing this effect is required by complete positivity. A central future direction is to go beyond specific systems and investigate the existence of the stationary state for general nonholonomic systems.

\section*{Acknowledgements}
S.G. and D.S. acknowledge support from NSF CAREER Grant No. DMR-1944967. S.G. also thanks the KITP Fellows Program at the Kavli Institute for Theoretical Physics, supported by NSF Grant No. PHY-2309135, where part of this work was carried out. SG would like to thank V. P. Nair, Alexios Polychronakos, and Alexander Abanov for useful discussions. 

	\bibliographystyle{unsrt}
	\bibliography{nonholonomic}
    
	\appendix
    	\setcounter{secnumdepth}{2}
        
		\section{Wigner transform of the master equation}\label{App wigner}

    The Wigner transform $\mathcal{W}[\hat{A}]\equiv A$ of an operator $\hat{A}$ involves taking a matrix element between position eigenstates $\langle z_+ | \hat{A} | z_-\rangle$, and then Fourier transforming in the relative coordinate $z_q=z_+-z_-$. This produces a function of the average position $z= (z_++z_-)/2$ and the Fourier momentum $p_z$,
    \begin{align}
    	A(z, p_z)=\int dz_q e^{-\frac{i}{\hbar}p_z z_q}\left\langle z+\frac{z_q}{2}\right|\hat{A}\left| z-\frac{z_q}{2}\right\rangle.
    \end{align}
    
    In the example of the skater system, there are three position coordinates $x, y, \Phi$, and we Wigner transform in each. The coordinate difference $\Phi_q$ has periodicity of $4\pi$, implying that the Fourier conjugate $p_\Phi = \hbar n$ with $n\in \frac{1}{2}\mathbb{Z}$. The argument of the Wigner function $\Phi$ has periodicity $2\pi$, and there is an identification $(\Phi+\pi, \Phi_q)\sim (\Phi, \Phi_q+2\pi)$. This implies the parity relation
    \begin{align}
    	A(\Phi+\pi, p_\Phi)=(-1)^{\frac{2p_\Phi}{\hbar}}A(\Phi, p_\Phi).
    \end{align} 
     In practice, we present our results in terms of a partial Wigner transform which uses the continuous $\Phi_q$ rather than $p_\Phi$. Some alternative approaches to the Wigner transform of a compact variable appear in, e.g., \cite{Mukunda1979,Kastrup2016}. 
    
     The Hamiltonian drift due to $\hat{p}_x^2+\hat{p}_y^2$, $\alpha \hat{y}$, and the $[\hat{A}_a, [\hat{A}_a, \hat{\rho}]]$ diffusion term in \eqref{Dissipator} do not involve the compact variable in an essential way,
     \begin{gather}
     	-\frac{1}{i\hbar}\frac{1}{2m}\mathcal{W}\left[\left[\hat{\rho}, \left(\hat{p}^2_x+\hat{p}^2_y\right)\right]\right]=-\frac{p_x}{m}\partial_x\rho-\frac{p_y}{m}\partial_y\rho,\\
     	-\frac{\alpha}{i\hbar}\mathcal{W}\left[\left[\hat{\rho}, \hat{y}\right]\right]=\alpha \partial_{p_y}\rho,\\
     	-\frac{\gamma}{2\hbar}\mathcal{W}\left[[\hat{A}_a, [\hat{A}_a, \hat{\rho}]]\right]=\frac{\gamma\hbar D}{2}\left(\partial_{p_x}^2+\partial_{p_y}^2\right)\rho.
     \end{gather}
     The partial Wigner transform of the drift due to $\hat{p}_\Phi^2$ is
     \begin{align}
     	-\frac{1}{i\hbar}\frac{1}{2I_0}\mathcal{W}\left[\left[\hat{\rho}, \hat{p}^2_\Phi\right]\right]=-\frac{1}{I_0}\left(-i\hbar\partial_{\Phi_q}\right)\partial_\Phi \rho.
     \end{align} 
    After Fourier transforming, $-i\hbar\partial_{\Phi_q}$ becomes $p_\Phi$, as one would anticipate.
     
    The remaining terms in the dissipator \eqref{Dissipator} are more involved,
 
    \begin{widetext}
\begin{gather}
\frac{\gamma}{i\hbar}\mathcal{W}\left[[\hat{A}_a,\{\hat{\rho},\hat{B}_a\}]\right]=	2\gamma\left(\cos^2\left(\frac{\Phi_q}{2}\right)\partial_{p_s} \left(p_s \rho\right)+\sin^2\left(\frac{\Phi_q}{2}\right)\partial_{p_r} \left(p_r\rho\right)+\frac{i\hbar}{4}\sin\Phi_q\left(\partial_{p_r}\partial_s + \partial_{p_s}\partial_r\right) \rho\right),\label{Wigner diss drift}
\end{gather}
\vspace{-12pt}
\begin{multline}
	-\frac{\gamma}{2\hbar}\mathcal{W}\left[[\hat{B}_a, [\hat{B}_a, \hat{\rho}]]\right]=\frac{\gamma }{D}\left(\frac{\hbar}{2}\cos^4\left(\frac{\Phi_q}{2}\right)\partial_s^2\rho-i\sin\Phi_q \cos^2\left(\frac{\Phi_q}{2}\right)p_r\partial_s \rho- \frac{\sin^2\Phi_q}{2\hbar}\left(p_r^2+p_s^2\right)\rho\right. \\
	\left.-i\sin\Phi_q \sin^2\left(\frac{\Phi_q}{2}\right)p_s\partial_r \rho+\frac{\hbar}{2}\sin^4\left(\frac{\Phi_q}{2}\right)\partial_r^2\rho\right).\label{Wigner B diff}
\end{multline}
    \end{widetext}
    Upon Fourier transforming in $\Phi_q$, the trigonometric factors become
    finite-difference operators on the $p_\Phi$ lattice. For example, $\sin\Phi_q\,f(\Phi_q)
    	\rightarrow
    	i\hbar\left(\frac{f(p_\Phi+\hbar)-f(p_\Phi-\hbar)}{2\hbar}\right).$
    	
    When $\rho(p_\Phi)$ varies slowly on the scale $\hbar$, these finite differences
    may be expanded in derivatives with respect to $p_\Phi$.\footnote{The integer and half-integer sublattices are interpolated separately as smooth functions of $p_\Phi$. The resulting functions of $(\Phi,p_\Phi)$ are respectively even and odd under $\Phi\to\Phi+\pi$. The finite-difference operators only involve values on the same sublattice.} To leading order, 	$\sin\Phi_q \rightarrow i\hbar\partial_{p_\Phi}+\order{\hbar^3},$ and $\sin^2\left(\frac{\Phi_q}{2}\right) \rightarrow -\frac{\hbar^2}{4}\partial^2_{p_\Phi}+\order{\hbar^4}.$
    
    Then the dissipative drift \eqref{Wigner diss drift} becomes $2\gamma \partial_{p_s}(p_s\rho)+\order{\hbar^2}$, and the  diffusion terms \eqref{Wigner B diff} become $\frac{\hbar\gamma}{2D}\mathcal{Q}\rho+\order{\hbar^3}$, leading to the Fokker-Planck equation \eqref{Master eq skater}.

	\section{More on the Suslov system}\label{App suslov}
	
	The configuration space of the Suslov system is the Lie group $SO(3)$ or its double cover $SU(2)$. Points in $SU(2)$ are parameterized by three angles $0\leq \theta\leq \pi, 0\leq\phi<2\pi, 0\leq \psi <4\pi$.

	Angular derivatives are taken with the left-invariant vector fields $X_a$,
	\begin{gather}
		X_1 = \sin\psi \partial_\theta + \cot \theta \cos\psi \partial_\psi - \csc \theta \cos\psi  \partial_\phi,\\\quad X_2 = \cos\psi \partial_\theta - \cot \theta \sin\psi \partial_\psi + \csc\theta \sin\psi  \partial_\phi,\\
		X_3 = \partial_\psi.
	\end{gather}
	
	The rotation matrices $R\in SO(3)$ are explicitly,
\begin{widetext}
	\begin{align}
	R = \left(\begin{array}{ccc}
		\cos\phi\cos\theta \cos\psi - \sin\phi\sin\psi & -\cos\phi\cos\theta \sin\psi - \sin\phi\cos\psi & \cos\phi\sin\theta \\
		\sin\phi\cos\theta \cos\psi +\cos\phi\sin\psi	& -\sin\phi\cos\theta \sin\psi +\cos\phi\cos\psi& \sin\phi\sin\theta\\
		-\sin\theta\cos\psi	&\sin\theta\sin\psi & \cos\theta
	\end{array}\right).\label{R matrix}
\end{align}
\end{widetext}

Besides the Lindblad operators $L_a$ in \eqref{Lindblad Suslov}, there is also an alternative set of Lindblad operators $\tilde{L}_a$ with the same classical drift but different diffusion terms
\begin{align}
		{L}_a = \sqrt{\kappa}\left(R_{a3}-i\frac{\mu}{2\kappa}\{R_{a2}, \omega_1\}\right),\nonumber\\
	\tilde{L}_a = \sqrt{\kappa}\left(R_{a2}+i\frac{\mu}{2\kappa}\{R_{a3}, \omega_1\}\right).\label{Lindblad Suslov tilded}
\end{align}
The master equation associated with either set of Lindblad operators is
\begin{gather}
	\dot{\rho}=\mathcal{L}_{SO(3)}\rho+\frac{\hbar \gamma \kappa}{2}\left(\partial_{j_1}^2+\partial_{j_n}^2\right)\rho + \frac{\hbar\gamma \mu^2}{2\kappa}\mathcal{Q}\rho,\\
	\mathcal{Q}\rho \equiv V^2 \rho + 2w_a\partial_{j_a}V\rho +\partial_{j_a}\left(Q_{ab}\partial_{j_b}\rho\right),\label{calQ Suslov}\\
	V\equiv -I^{-1}_{1a}X_a, \qquad w_a\equiv \epsilon_{abc}I^{-1}_{1b}j_c,\\
	Q_{ab}\equiv \omega_1^2\left(\delta_{ab}-\delta_{an}\delta_{bn}\right)+w_a w_b.
\end{gather}
The index $n=2$ or $3$ for $L$ and $\tilde{L}$, respectively, and as usual a weighted sum of both sets of operators is possible.  

Note that both $L_a$ and $\tilde{L}_a$ are distinct from the closely related Lindblad operators $A_a^{(1)}$ appearing in \cite{Stickler2018Rotational} for the purpose of representing a thermal bath. In our notation, and with a definite choice of operator ordering,
\begin{align}
	A^{(1)}_a=R_{a1}-i\frac{\mu}{2\kappa}\left(\{R_{a3}, \omega_2\}-\{R_{a2}, \omega_3\}\right).\label{sticklerEtAl lindblad}
\end{align}
The diffusion constant in $A$ is related to the inverse temperature through the relation $\mu/\kappa = \hbar\beta/4$. 
	\section{The $a\neq 0$ Chaplygin sleigh}\label{App chaplygin}
	
	The Chaplygin sleigh \cite{Chaplygin1912ReducingMultiplier, Caratheodory1933}, which includes the skater system as a special case, may be understood abstractly as a Suslov system on the Lie group $SE(2)$ of translations and rotations on the Euclidean plane.
	
	To understand $SE(2)$ as a limiting case of our previous results for the group $SO(3)$, introduce the auxiliary length scale $l$ and the coordinates
	\begin{align}
	x = l\cos\phi\sin\theta,\quad y = l\sin\phi\sin\theta,\quad \Phi = \psi+\phi.
	\end{align}
	Then expanding the matrix $R$ to leading order in $l^{-1}$,
	\begin{align}
		R = \left(\begin{array}{ccc}
			\cos\Phi & -\sin\Phi & \frac{x}{l} \\
			\sin\Phi	& \cos\Phi & \frac{y}{l} \\
			-\cos\Phi\frac{x}{l}  -\sin\Phi\frac{y}{l} 	&\sin\Phi\frac{x}{l} -\cos\Phi\frac{y}{l}  & 1
		\end{array}\right).
	\end{align}
	If only leading order terms in $l^{-1}$ are retained under matrix multiplication these matrices form a concrete realization of $SE(2)$.
	
	The Chaplygin sleigh Hamiltonian may be written in the form $H=\frac{1}{2}j_a I^{-1}_{ab}j_b$ if we understand
	\begin{gather}
		j_1 = -lp_s,\quad j_2=l p_r, \quad j_3 = p_\Phi,\label{j decompactification}\\
		I^{-1} = \frac{1}{mI_0\,l^2}\left(\begin{array}{ccc}
		I_0 + ma^2 & 0 & ma\,l \\
			0	& I_0 & 0 \\
		 ma\,l	&0 & m\,l^2
		\end{array}\right).\label{Moment of inertial Chaplygin}
	\end{gather}
	
	Taking the large $l$ limit with $D = \kappa l^{-2}$ and $m=\mu l^{-2}$ held fixed, the  Lindblad operators $L_1, L_2$ for the $SO(3)$ Suslov system  \eqref{Lindblad Suslov} simply reduce to the previous Lindblad operators for Chaplygin in \eqref{Lindblad operators Chaplygin}, with $v_s$ given by
	\begin{align}
		mv_s = \left(1+\frac{ma^2}{I_0}\right)p_s-\frac{ma}{I_0}p_\Phi.
	\end{align} To leading order the third operator $L_3$ only serves to provide a drift which may be absorbed into the Hamiltonian renormalization.
	
	Note in passing that if we instead take the limiting case of the operators $\tilde{L}_a$ in \eqref{Lindblad Suslov tilded}, we get an alternative set of three Lindblad operators: $-\sqrt{\kappa}\sin\Phi-\frac{im}{2\sqrt{\kappa}}\{x,v_s\}, \sqrt{\kappa}\cos\Phi-\frac{im}{2\sqrt{\kappa}}\{y,v_s\}, -\sqrt{D}s-\frac{im}{\sqrt{D}}v_s.$ These produce the correct classical drift, but the diffusion terms are not translation invariant.
	
	The master equation for the Chaplygin sleigh may be found directly from $L_1, L_2$ in \eqref{Lindblad operators Chaplygin} or by taking the large $l$ limit of the master equation for the Suslov system,
	\begin{align}
\dot{\rho}=\mathcal{L}\rho+\frac{\hbar\gamma D}{2}\left(\partial_{p_r}^2+\partial_{p_s}^2\right)\rho+\frac{\hbar\gamma}{2D}\mathcal{Q}\rho.
	\end{align}
	As before, $\mathcal{L}\rho = -[\rho, H]_P + 2\gamma \partial_{p_s}\left(mv_s \rho\right)$ with the $a\neq 0$ expressions for $H$ and $mv_s$ given above.
	
	The diffusion operator $\mathcal{Q}$ takes the same form as in the Suslov system \eqref{calQ Suslov}, but with definitions
	\begin{gather}
		V=\left(1+\frac{ma^2}{I_0}\right)\partial_s-\frac{ma}{I_0}\partial_\Phi,\\
		w=\left(-\frac{ma}{I_0}p_r, -\frac{ma}{I_0}p_s, \left(1+\frac{ma^2}{I_0}\right)p_r\right),\\
		Q_{ab}=w_a w_b + \left(mv_s\right)^2\delta_{a3}\delta_{b3}.
	\end{gather}
	In the $a=0$ limit this indeed reduces to $\mathcal{Q}$ for the skater system \eqref{Diff term}.

\section{More general diffusion terms}\label{App general diffusion}
        
The set of two Lindblad operators in \eqref{Lindblad operators Chaplygin} is not the only choice that reduces to the proper damping term \eqref{Liouville Chaplygin} in the classical limit. Instead of beginning with a choice of non-Hermitian Lindblad operators $L_a$, we may begin with an expanded basis of Hermitian operators $F_A$, and write the dissipator as
\begin{align}
	\mathcal{D}\rho=\frac{1}{\hbar}\sum_{AB}K^{AB}\left(F_A \rho F_B-\frac{1}{2}\left\{F_BF_A, \rho\right\}\right),
\end{align} 
where the Kossakowski matrix $K$ is Hermitian and is required to have non-negative eigenvalues \cite{Gorini1976}. Rather than choosing the set of $L_a$ in advance, we adjust the values of $K$, and upon diagonalizing the matrix we recover our earlier form for the dissipator \eqref{Master eq}.

It is enlightening to decompose $K=K_R+iK_I$ with $K_R, K_I$ symmetric and antisymmetric, respectively.
\begin{align}
	\mathcal{D}\rho=\frac{1}{2\hbar}\sum_{AB}\left(K_R^{AB}[F_A,[\rho,F_B]]+iK_I^{AB}[F_A,\{\rho, F_B\}]\right).\label{Kossakowski decomposition}
\end{align} 
So the real part is associated with diffusion terms and the imaginary part is associated with the drift terms.

An expanded operator basis for the skater case is given by the eight elements 
\begin{multline}
F_A = \left(x,\, y, \,p_x ,\, p_y ,\,\frac{1}{2}\cos(2\Phi) \,p_x,\,\frac{1}{2}\sin(2\Phi) \,p_x,\right.\\
\left.\frac{1}{2}\cos(2\Phi) \,p_y,\,\frac{1}{2}\sin(2\Phi) \,p_y\right).
\end{multline}
This is sufficiently general to allow for damping in the $p_s$ direction parameterized by $\gamma_s$, as well as damping in the perpendicular $p_r$ direction, with coefficient $\gamma_r$. The imaginary part $K_I$ will be fixed by these damping terms.

There is more freedom to choose the real part $K_R$, but it is restricted by symmetry. Consider the generator $J= x p_y - y p_x + p_\Phi$. The dissipator respects symmetry if $[\mathcal{D}\rho, J]= \mathcal{D}\left([\rho, J]\right).$
If we act with the symmetry generator on the basis elements $[F_A, J]= F_{A'}M^{A'}_{\quad A},$
then symmetry implies the matrix relation $M K+K M^{T}=0.$

\begin{widetext}
	Subject to these requirements, the most general Hermitian $K$ matrix in this basis involves ten parameters $c_i$,
\begin{align}
	K &= \left(\begin{array}{cc|cc|cccc}
		c_1 & 0 & c_7 - i\gamma & c_8 & c_9 + i\delta & c_{10} & -c_{10} & c_9+i\delta \\
		& c_1 & -c_8 & c_7 - i\gamma & -c_{10} & c_9+i\delta & -c_9-i\delta& -c_{10} \\ \hline
		& &	c_2 & 0 & c_5 & c_6 & -c_6 & c_5\\
		& & & c_2 & -c_6 & c_5 & -c_5 & -c_6\\ \hline
		& & & & c_3 & 0 & 0 & c_{4}\\
		& & & & & c_3 & -c_{4 }& 0\\
		& & & & & & c_3 & 0\\
		& & & & & & & c_3
	\end{array}\right),\qquad \gamma\equiv \frac{\gamma_s+\gamma_r}{2},\quad \delta\equiv \gamma_s-\gamma_r.
\end{align}	
\end{widetext}

If both $\gamma$ and $\delta$ are non-zero, then the requirement that K have nonnegative eigenvalues implies $c_1, c_2, c_3>0$. In the Fokker-Planck approximation, these coefficients correspond to the diffusion terms
\begin{multline}
	\frac{\hbar}{2}\left[c_1\left(\partial_{p_r}^2+\partial_{p_s}^2\right)+c_2\left(\partial_s^2+\partial_r^2\right)\right.\\
	\left.+c_3\left(
	(p_r^2+p_s^2)\partial_{p_\Phi}^2
	+\frac{1}{4}(\partial_s^2+\partial_r^2)
	\right)\right]\rho.
\end{multline}
If all the other coefficients $c=0$, non-negativity implies the inequality
\begin{align}
	c_1 c_2\geq \gamma^2+2\delta^2\frac{c_2}{c_3}. \label{c123 inequality}
\end{align}
In the special case that $\delta=0$ we are allowed to take $c_3=0$. This makes sense since the case $\gamma_s=\gamma_r$ is equivalent to the case of ordinary isotropic dissipation in two dimensions. But as long as we have $\gamma_s\neq \gamma_r$, complete positivity requires $c_3>0$, and thus the $(p_r^2+p_s^2)\partial^2_{p_\Phi}$ diffusion term is unavoidable. This conclusion is unchanged by enlarging the operator basis, since the Kossakowski matrix above remains a principal submatrix of the enlarged matrix.

Note that if the only non-zero parameters are $c_1, c_2, c_3$ and the inequality \eqref{c123 inequality} is saturated, this actually corresponds to $6$ distinct Lindblad operators, whereas we needed only $2$ operators $L_a$ in the main text \eqref{Lindblad operators Chaplygin}. The earlier pair of Lindblad operators involves the choice $c_1=\gamma_s D$, $4c_2=c_3 = c_4= -2c_5=\gamma_s/D$. The additional $c_4, c_5$ diffusion terms take the form
\begin{multline}
	\frac{\hbar}{2}\left[c_4\left(
	p_r\,\partial_s\partial_{p_\Phi}
	-p_s\,\partial_r\partial_{p_\Phi}
	\right)\right.\\
	\left.+c_5 \left(
	\partial_r^2-\partial_s^2
	-2p_r\,\partial_s\partial_{p_\Phi}
	-2p_s\,\partial_r\partial_{p_\Phi}
	\right)\right]\rho.
\end{multline}

The remaining parameters lead to mixed derivatives $\partial_s \partial_r$ (for $c_6$), and mixed derivatives in position and translational momentum ($c_{7}$ through $c_{10}$), together with requisite terms involving derivatives in $p_\Phi$. The leading-order terms may be calculated by replacing commutators with Poisson brackets in \eqref{Kossakowski decomposition} and we omit them here.

The operator basis may of course be enlarged. In particular, including the operator $p_\Phi$ would allow for diffusion in the $\Phi$ coordinate, and including the set of four operators $\cos\Phi, \sin\Phi, \frac{1}{2}\left\{\cos\Phi, p_\Phi\right\}, \frac{1}{2}\left\{\sin\Phi, p_\Phi\right\}$ would allow treatment of the $a\neq 0$ Chaplygin case, as well as damping in the $p_\Phi$ direction.

On this last point note that the set of two Lindblad operators \cite{Stickler2018Rotational}
\begin{multline}
	L_{1}= \sqrt{\kappa}\cos\Phi-\frac{i}{2\sqrt{\kappa}}\left\{\sin\Phi, p_\Phi\right\},\\L_{2}= \sqrt{\kappa}\sin\Phi+\frac{i}{2\sqrt{\kappa}}\left\{\cos\Phi, p_\Phi\right\},
\end{multline}
lead to damping of the angular momentum $p_\Phi$. These operators follow from the decompactification limit \eqref{j decompactification} of the Lindblad operators for the Suslov system $\tilde{L}_a^{(3)}=\sqrt{\kappa}\left(R_{a1}+i\frac{1}{2\kappa}\{R_{a2}, j_3\}\right)$ with a diagonal moment of inertia tensor and damping in the $j_3=\mu \omega_3$ direction.

		\section{Numerical simulation}\label{App simulation}
	
For the skater system, Suslov system, and Chaplygin sleigh, we reduce the master equation to a Fokker-Planck equation for a probability distribution $f$ as a function of $t$ and phase space coordinates $x_i$. Schematically,
\begin{align}
	\dot{f}= -\partial_{x_i}(A_i(x)\,f)+\frac{1}{2}\partial_{x_i}\partial_{x_j}\left(Q_{ij}(x)f\right).\label{Fokker Planck generic}
\end{align}
As is well known (see e.g. \cite{Gardiner1985}), the Fokker-Planck equation may be simulated by sampling an associated Langevin equation, which is an It\^{o} stochastic differential equation of the form
\begin{align}
	d{x}_i= A_i(x)dt+B_{i\nu}(x)dW^\nu,
\end{align}
with $B$ chosen so that $Q_{ij}= \sum_\nu B_{i\nu}B_{j\nu}.$

We may decompose the drift into two terms $A_i(x) = a_i(x)+\gamma L_i^{\,\,j}x_j$ where $\gamma$ is a large parameter. To treat these stiff linear terms in a discrete time approximation, we use a first-order exponential time differencing (ETD) scheme \cite{CoxMatthews2002}. This involves factorizing $x(t)= e^{\gamma L t}u(t)$, and integrating the equation for $u$ over the time step $\Delta t$, making the first-order approximation that $a(t)$ and $B(t)$ are constant over the time step.

In the case of the skater system \eqref{Master eq skater}, we simulate the full Langevin equation involving both position and momentum variables $\Phi, p_\Phi, r, p_r, s, p_s$. Since the deterministic evolution with finite damping $\gamma$ is calculable in closed form, we use the exact solution $x_{det}(t+\Delta t | x(t))$ given initial conditions at $t$ instead of taking the non-dissipative drift $a$ to be constant over time step $\Delta t$.

In the absence of an ETD correction to the noise, the finite difference step would look like
\begin{align}
	x(t+\Delta t)= x_{det}(t+\Delta t| x(t)) + \sqrt{\hbar\gamma \Delta t}B(x(t))\xi.
\end{align}
The $B_{i\nu}$ matrix is chosen to multiply four independent Gaussian noises $\xi^\nu$ with unit variance, and it has non-zero components $B_{p_s\, 1}=B_{p_r\,2}=1$, $B_{p_\Phi\,3}=p_s$, $B_{p_\Phi\,4}=p_r$, $B_{s\,4}=1$.

Since the damping only acts in the $p_s$ direction, the only effect of the ETD scheme in this case is to replace $\Delta t$  in the equation for $p_s$ by the effective time step $\Delta t_{p_s}$
\begin{align}
	\Delta t_{p_s}= \frac{1-e^{-4\gamma \Delta t}}{4\gamma}.
\end{align}
If $\gamma \Delta t\ll 1$ this reduces to the ordinary time step, but this correction allows us to use a $\Delta t $ which is comparable to the dissipative time $\gamma^{-1}$ but still small compared to other time scales in the system. The results for $\gamma=100, \hbar=10^{-7}, \Delta t=.01$ (in dimensionless units) are plotted for $\alpha=2/3$ in Fig.~\ref{fig:skaterXY} and $\alpha=0$ in Fig.~\ref{fig:skaterVar}. Initial mean values of a minimum uncertainty wave packet are set as $p_r=p_\Phi=1, p_s=0, \Phi = -\pi/2$.

In the case of the Suslov system \eqref{Master equation Suslov}, we integrate over the Euler angles to get a Fokker-Planck equation for the marginal distribution $f$ as a function of time and the momentum $j_a$ alone. Following \eqref{calQ Suslov}, the diffusive terms in the Fokker-Planck equation are written in the form $\partial_{j_a}\left(Q_{ab}\partial_{j_b} f\right)$, and thus upon putting the Fokker-Planck equation in the form \eqref{Fokker Planck generic} we generate an additional \emph{diffusive} drift term 
\begin{align}A_{diff,a}=\frac{\hbar\gamma}{2}\partial_{j_b}Q_{ab}.\end{align}

For the purposes of an ETD scheme, both the Hamiltonian and diffusive drift terms are grouped into the drift $a(j)$, and the stiff damping term
\begin{align}
	\gamma L_{ab}= -2\gamma \delta_{a1}I^{-1}_{b1},
\end{align}
is exponentiated to
\begin{gather}
	e^{\gamma L t}= I-P+e^{-2\gamma I^{-1}_{11}t}P,\quad P_{ ab}=\frac{\delta_{a1}I^{-1}_{b1}}{I^{-1}_{11}}.\label{expL}
\end{gather}
Then the ETD approximation to the continuous time stochastic differential equation is
	\begin{multline}
	    	j(t+\Delta t) = e^{\gamma L \Delta t}j(t)\\+\left(\int_0^{\Delta t}e^{\gamma L(\Delta t- s)}ds\right)a(j(t))+\xi\left(j(t),\Delta t\right),
	\end{multline}
where $\xi$ is a Gaussian random variable with zero mean and covariance matrix $\tilde{Q}_{ab}=\langle \xi_a\xi_b \rangle$ given by
\begin{align}
\tilde{Q}=\int_0^{\Delta t} e^{\gamma Ls}Q e^{\gamma L^Ts}ds.
\end{align}
We carry out these integrals exactly using the explicit form of the matrix exponential \eqref{expL}.

Fig.~\ref{fig:suslovOmega} shows the results of a Langevin simulation of the Suslov system with $\gamma=1000, \hbar=10^{-6}, \Delta t=.01$ and moment of inertia tensor
\begin{gather}
I = \left(\begin{array}{ccc}
			1 & 1/4 & -1/2 \\
			1/4	& 1 & 0 \\
		-1/2	&0 & 2
		\end{array}\right).
	\end{gather}

	\section{Solving the truncated master equation}\label{App ornstein uhlenbeck}
        
	After truncating to the lowest order in the weak noise approximation (see e.g. \cite{Gardiner1985}), the master equation \eqref{Master eq weak noise} for the skater system with $\alpha=0$ takes the form of an Ornstein-Uhlenbeck equation with time-dependent matrix coefficients $A, Q$. Schematically,
	\begin{align}
		\dot{f}=-\partial_{z^i}\left(A^i_{\,\,j}z^j\,f\right)+\frac{1}{2}\partial_{z^i}\partial_{z^j}\left(Q^{ij}f\right),\label{Master eq OU}
	\end{align}
	where in our case, the coordinates $z$ are $(q_r, q_s, q_\Phi)$, and the drift matrix $A$ is composed of blocks $A_0$ and $\tilde{p}$,	\begin{gather}
		A(t)=	\left(\begin{array}{cc}
			A_0 & \tilde{p}(t)\\
			0 & 0
		\end{array}\right),\\ A_0 = \left(\begin{array}{cc}
			0& \omega \\
			-\omega & -2\gamma
		\end{array}\right),\quad \tilde{p}(t)=\left(\begin{array}{c}
			\bar{p}_s(t)\\
			-\bar{p}_r(t)
		\end{array}\right),
	\end{gather}
	and the diffusion matrix $Q$ is
	\begin{gather}
		Q(t)=	\left(\begin{array}{cc}
			I & 0\\
			0 & \bar{p}_r(t)^2+\bar{p}_s(t)^2
		\end{array}\right).
	\end{gather}
	The Ornstein-Uhlenbeck equation \eqref{Master eq OU} implies that the covariance matrix $\Sigma(t)$ satisfies	$\dot{\Sigma}= A\Sigma+\left(A\Sigma\right)^T+Q,$ and this has solution
	\begin{multline}
		\Sigma(t)=\Phi(t,0)\Sigma(0)\Phi(t,0)^T\\+\int^t_0 \Phi(t,s)Q(s)\Phi(t,s)^Tds,
	\end{multline}
	where $\Phi$ is a time-ordered exponential,
		\begin{gather}
		\Phi(t,s)=\mathcal{T}\exp\left(\int_s^t A(\tau)d\tau\right)=\left(\begin{array}{cc}
			e^{A_0(t-s)} & h(t,s)\\
			0 & 1
		\end{array}\right),\\
		h(t,s)=\int_s^td\tau  e^{A_0(t-\tau)}\tilde{p}(\tau).
	\end{gather}

	This can be calculated exactly given the functions $p=(\bar{p}_r, \bar{p}_s)^T$ which are classical solutions to the skater system with $\bar{p}_\Phi$ equal to the constant $\omega$. They satisfy equations of motion $\dot{p}= A_0 p$, with explicit solution $\bar{p}_r=C_+ e^{\lambda_+ t} + C_- e^{\lambda_- t}$ and $\bar{p}_s = \omega^{-1}\dot{\bar{p}}_r$. Here $\lambda_\pm$ are the eigenvalues of $A_0$,
	\begin{align}		
		\lambda_\pm = -\gamma\pm \sqrt{\gamma^2-\omega^2},
	\end{align}
	and $C_\pm$ are fixed by the initial conditions,
	\begin{align}		
		C_\pm = \pm \frac{\omega p_{s,0}-\lambda_{\mp}p_{r,0}}{\lambda_+-\lambda_-}.
	\end{align}
	
	Note that this solution gives another perspective on the variance relation \eqref{Variance nonholonomic} which was earlier derived from the metastable manifold perspective. Let us introduce the instantaneous left eigenvector of $A$,
	\begin{gather}
	v_-(t)=\left(-\frac{\omega}{\lambda_-},\, 1,\,\,c_-(t)\right)\approx \left(\frac{\omega}{2\gamma},\, 1,\,\,\frac{\bar{p}_r(t)}{2\gamma}\right),\\
	c_-(t)\equiv -\frac{\bar{p}_r(t)}{\lambda_-}
	-\frac{\omega\bar{p}_s(t)}{\lambda_-^2}.
\end{gather}
In the normalization of the $q$ variables, the variance relation \eqref{Variance nonholonomic} may be written,
\begin{align}
	v_-(t) \Sigma(t) v_-(t)^T= \frac{1}{4\gamma}+\order{\gamma^{-3}}.\label{Variance nonholonomic 2}
\end{align}
We will proceed to demonstrate this relation.

$v_-$ is approximately a left eigenvector of the matrix exponential
	\begin{align}
	v_-(t)\Phi(t,s)
	\approx e^{\lambda_-(t-s)}v_-(s),
\end{align}
where the correction term comes from an integral $\int_s^t d\tau\,
e^{\lambda_-(t-\tau)}\dot{c}_-(\tau)\sim \order{\gamma^{-3}}$.

Taking $t\gg \gamma^{-1}$, the initial value term in $v_- \Sigma v_-^T$ is exponentially damped,
\begin{align}
	v_-(t)\Sigma(t)v_-(t)^T &\approx \int^t_0 e^{2\lambda_-(t-s)}v_-(s)Q(s)v_-(s)^Tds\nonumber\\
	&\approx \frac{v_-(t)Q(t)v_-(t)^T}{-2\lambda_-}.
\end{align}
Now $-2\lambda_-=4\gamma\left(1+\order{\gamma^{-2}}\right)$, and the leading order term of the numerator is $1$, which comes from the second component of $v_-$, the other components contributing at $\order{\gamma^{-2}}$. This demonstrates \eqref{Variance nonholonomic 2}.

\end{document}